\documentclass[aps,prb,reprint,superscriptaddress]{revtex4-1}

\usepackage[pdftex]{graphicx}
\usepackage{color}
\usepackage{enumitem}
\usepackage{gensymb}
\usepackage{epstopdf}
\usepackage{bm}
\usepackage{amsmath,amsthm,amssymb}
\usepackage{dcolumn}

\begin{document}


\title{Universality of the helimagnetic transition in cubic chiral magnets: Small angle neutron scattering and neutron spin echo spectroscopy studies of Fe$_{1-x}$Co$_x$Si}



\author{L.J. Bannenberg}
\affiliation{Faculty of Applied Sciences, Delft University of Technology, Mekelweg 15, 2629 JB Delft, The Netherlands}
\author{K. Kakurai}
\affiliation{Neutron Science and Technology Center, CROSS Tokai, Ibaraki 319-1106, Japan}
\affiliation{RIKEN Center for Emergent Matter Science(CEMS), Wako 351-0198, Japan}
\author{P. Falus}
\affiliation{Institut Laue-Langevin, 71 Avenue des Martyrs, CS 20156 Grenoble, France}
\author{E. Leli\`{e}vre-Berna}
\affiliation{Institut Laue-Langevin, 71 Avenue des Martyrs, CS 20156 Grenoble, France}
\author{R.M. Dalgliesh}
\affiliation{ISIS, Rutherford Appleton Laboratory, STFC, OX11 0QX Didcot, United Kingdom}
\author{C.D. Dewhurst}
\affiliation{Institut Laue-Langevin, 71 Avenue des Martyrs, CS 20156 Grenoble, France}
\author{F. Qian}
\affiliation{Faculty of Applied Sciences, Delft University of Technology, Mekelweg 15, 2629 JB Delft, The Netherlands}
\author{Y. Onose}
\affiliation{Department of Basic Science, University of Tokyo, Tokyo, 153-8902, Japan}
\author{Y. Endoh}
\affiliation{RIKEN Center for Emergent Matter Science(CEMS), Wako 351-0198, Japan}
\author{Y. Tokura}
\affiliation{RIKEN Center for Emergent Matter Science(CEMS), Wako 351-0198, Japan}
\affiliation{Department of Applied Physics, University of Tokyo, Tokyo 113-8656, Japan}
\author{C. Pappas}
\affiliation{Faculty of Applied Sciences, Delft University of Technology, Mekelweg 15, 2629 JB Delft, The Netherlands}

\date{\today}

\begin{abstract}
We present a comprehensive Small Angle Neutron Scattering (SANS) and Neutron Spin Echo Spectroscopy (NSE) study of the structural and dynamical aspects of the helimagnetic transition in Fe$_{1-x}$Co$_x$Si with $x$ = 0.30. In contrast to the sharp transition observed in the archetype chiral magnet MnSi, the transition in Fe$_{1-x}$Co$_x$Si is gradual and long-range helimagnetic ordering coexists with short-range correlations over a wide temperature range. The dynamics are more complex than in MnSi and involve long relaxation times with a stretched exponential relaxation which persists even under magnetic field. These results in conjunction with an analysis of the hierarchy of the relevant length scales show that the helimagnetic transition in Fe$_{1-x}$Co$_x$Si differs substantially from the transition in MnSi and question the validity of a universal approach to  the helimagnetic transition in chiral magnets.
\end{abstract}


\maketitle

\section{\label{sec:level1} Introduction}
Cubic helimagnets such as MnSi, FeGe, Cu$_2$OSeO$_3$ and Fe$_{1-x}$Co$_x$Si attract a great amount of attention due to  the observation of chiral skyrmions and their lattices.\cite{muhlbauer2009,FeGe_Lorenz_TEM,wilhelm2011,seki2012observation,yu2010,munzer2010} These chiral skyrmions have dimensions significantly larger than the lattice constant, are topologically protected and may have applications in spintronics and novel devices for information storage.\cite{nagaosa2013,fert2013,romming2013}

In these chiral magnets, a long-range helimagnetic order of the magnetic moments exists at zero field below the critical temperature $T_C$. The helimagnetic ordering is the result of the competition between three hierarchically-ordered interactions,\citep{bak1980} of which the strongest is the ferromagnetic exchange interaction favoring parallel spin alignment. The twist of the spins is induced by the weaker Dzyaloshinsky-Moriya (DM) interaction that results from the absence of a center of symmetry of the crystallographic structure.\citep{D,M} The propagation vector of the resulting helical arrangement of the magnetic moments is fixed by anisotropy. If a magnetic field is applied that is sufficiently strong to overcome the anisotropy, it aligns the helices along its direction and induces the so called conical phase. Within this conical phase, skyrmion lattice correlations are stable in a small pocket just below $T_C$,\cite{muhlbauer2009, munzer2010, wilhelm2011, seki2012observation} and metastable in a much larger region of the magnetic phase diagram.\cite{munzer2010,bannenberg2016,oike2016} 

In helimagnets, theory predicts at zero magnetic field a first order transition to the helimagnetic state.\citep{bak1980}  In the archetype chiral magnet MnSi this is indeed confirmed by sharp anomalies of the thermal expansion,\citep{matsunaga1982,stishov2007} heat capacity,\citep{stishov2007,bauer2013} and ultrasound absorption\citep{kusaka1976,petrova2009} at $T_C$. In this system, strong fluctuating correlations build up just above $T_C$ and thus precede the first order phase transition. These correlations show up as a ring of intensive diffuse neutron scattering spreading over a surface with radius $\tau$ = 2$\pi$/$\ell$,\citep{grigoriev2005} where $\ell$ denotes the pitch of the helix.  The origin of this precursor phase remains subject to debate. Based on the observations that these correlations are totally chiral up to $\sim$ $T_C$ + 1~K, it was suggested  that this scattering might emanate from a chiral spin liquid phase which would be the magnetic equivalent of the blue phase observed in liquid crystals.\citep{pappas2009} It has also been argued that the correlations drive the transition to first order as suggested by Brazovskii in a theory originally developed for liquid crystals.\citep{brazovskii1975,janoschek2013} This approach provides a good description of the temperature dependence of the susceptibility and correlation length, but does not explain all intriguing features of the precursor phase and the transition in MnSi.\citep{sidorov2014,stishov2016} Studies of the helimagnetic transition to other cubic chiral magnets are scarce. In fact only Cu$_2$OSeO$_3$ has been studied by a critical scaling analysis and in this case, the Brazovskii approach was found to be less conclusive than for MnSi.\citep{zivkovic2014,sidorov2014}

In this work, we address the open question of the helimagnetic transition in the semiconductor Fe$_{1-x}$Co$_x$Si and at the same time of the theoretically expected universality of the helimagnetic transition in cubic chiral magnets.\cite{janoschek2013} Fe$_{1-x}$Co$_x$Si is of particular interest as important physical properties can be altered by tuning the chemical substitution which changes both the sign and the magnitude of the DM-interaction. The helical order is stabilized over a wide range of concentrations of 0.05 $<$ $x$ $<$ 0.8.\cite{beille1981,beille1983,motokawa1987} By changing the concentrations, $T_C$ changes from a few Kelvin to 50 K and the pitch $\ell$ from $\sim$30 nm to $\sim$200 nm.\citep{beille1983,grigoriev2009} Furthermore, the sign of the chirality alters from left to right-handed at $x$ = 0.65.\citep{siegfried2015} 

The specific composition of the sample used in this work, Fe$_{0.7}$Co$_{0.3}$Si, has a $T_C$ of approximately 43~K and a pitch of $\ell$ $\sim$40~nm. We present the results of Small Angle Neutron Scattering (SANS) measurements that provide structural information on the magnetic correlations as well as SANS in combination with polarization analysis to determine the degree of magnetic chirality. These measurements are complemented by the investigation of the associated dynamics by Neutron Spin Echo Spectroscopy (NSE). The combined experimental findings show that the helimagnetic transition in Fe$_{0.7}$Co$_{0.3}$Si is gradual and involves slow and complicated dynamics and is as such quantitatively different from the transition in MnSi, which challenges the validity of an universal approach to the helimagnetic transition for chiral magnets.

\section{\label{sec:level2} Experimental Details}
The measurements were performed with the Fe$_{0.7}$Co$_{0.3}$Si single crystal ($\sim$0.1~cm$^3$) that was  used for previous neutron scattering studies\cite{takeda2009,bannenberg2016} and originates from the same batch as the sample for the ac susceptibility measurements.\cite{bannenberg2016squid} The sample was oriented with the [$\bar{1}$10] axis vertical for all experiments.

SANS measurements were performed on the instruments D33 at the Institute Laue Langevin and LARMOR at ISIS. At D33, the monochromatic neutron beam had an incident wavelength of $\lambda$ = 0.6~nm with $\Delta\lambda / \lambda = 10\%$ and the magnetic field $\vec{B}$ was applied along $\vec{k_i}$, the wavevector of the incoming neutron beam. Complementary measurements with $\vec{B} \perp \vec{k}_i$ were performed on the time-of-flight SANS instrument LARMOR at the ISIS neutron spallation source where neutrons with wavelengths of 0.8$\, \leq \lambda  \leq \,$1.6~nm where used. The SANS patterns were normalized to standard monitor counts and background corrected using a  measurement at 60~K, a temperature which corresponds to $\sim$1.5 $T_C$. Measurements with $\vec{B} \perp \vec{k}_i$ were performed after either Zero Field Cooling (ZFC) the sample or by Field Cooling (FC) through $T_C$. As the results did not depend on the specific magnetic history and the specific protocol, most measurements were recorded with a ZFC protocol.

Neutron Spin Echo, SANS with polarization analysis and spherical polarimetry were performed on the NSE spectrometer IN15 at the Institut Laue Langevin using  a polarized neutron beam with a polarization of 95\% and a monochromatization of $\Delta\lambda / \lambda = 15\%$. At zero field, both the paramagnetic NSE, SANS with polarization analysis and spherical polarimetry measurement were performed with $\lambda$ = 0.8~nm. The measurements under magnetic field were performed with $\lambda$ = 1.2~nm and in the ferromagnetic NSE configuration.\cite{Farago:1986vl, Pappas:2008fb} For these measurements the magnetic field was applied perpendicular to the incident neutron beam ($\vec{B}\perp \vec{k}_i$), a configuration where the chiral scattering of the sample does not depolarize the scattered neutron beam.\cite{pappas2011} All NSE spectra were averaged over the entire detector and a background correction was performed when required using a high temperature measurement at 60~K.
 
\begin{figure*}
\begin{center}
\includegraphics[width= 1 \textwidth]{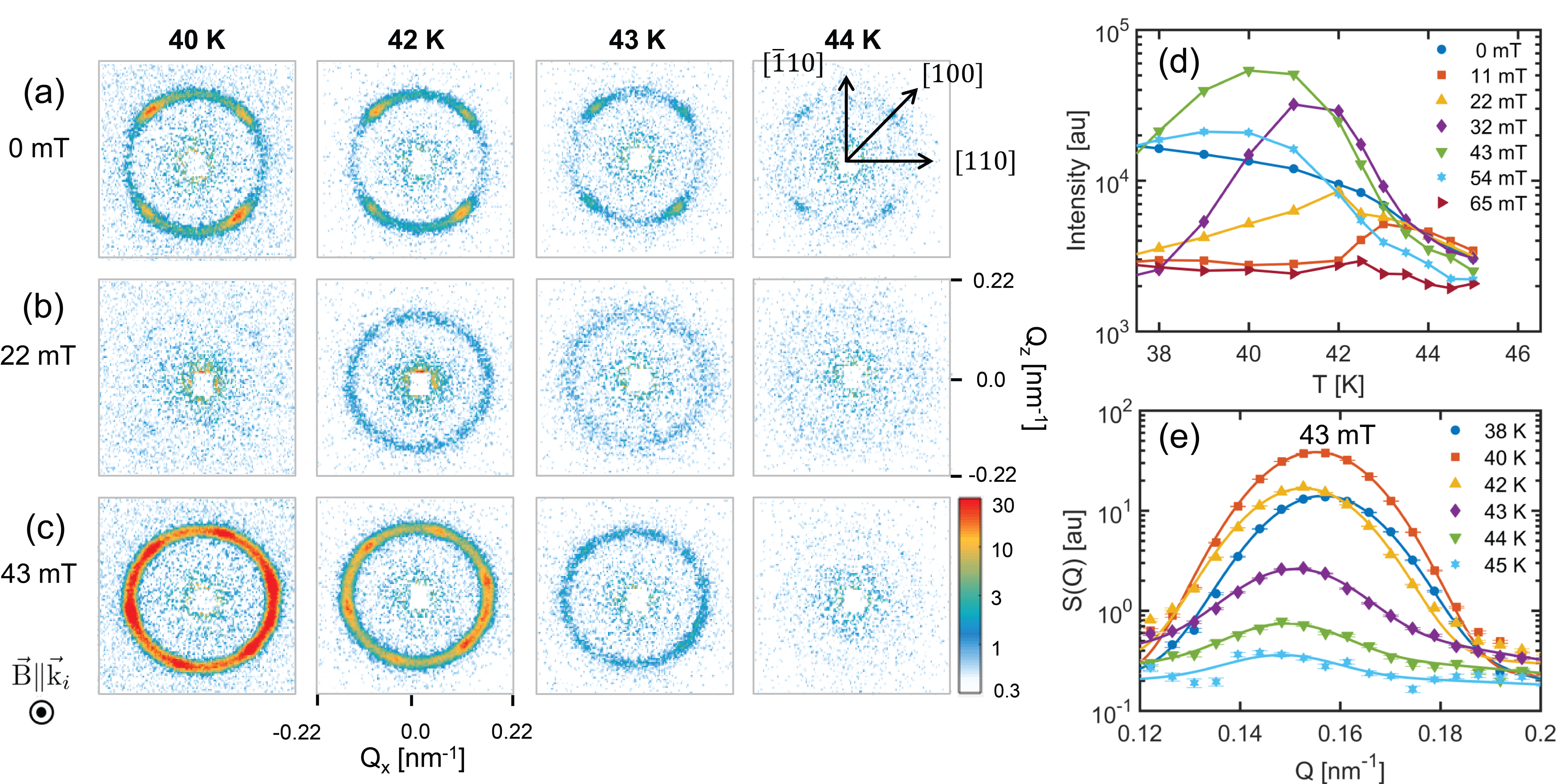}
\caption{SANS results obtained at D33 by applying the magnetic field along the incident neutron beam ($\vec{B} \parallel  \vec{k}_i$). Characteristic patterns are shown for $B$ = 0 mT, 22 mT and 43 mT in panels (a) - (c). Panel (d) displays the temperature dependence of the total scattered intensity for selected magnetic fields. Panel (e) shows $S(Q)$ in arbitrary units, deduced by  radially averaging the scattered intensity at a magnetic field of 43~mT, for the temperatures indicated. The solid lines indicate the best fits of  eq. \ref{LorGauss} to the data.}
\label{SANS_D33}
\end{center}
\end{figure*}

\begin{figure*}
\begin{center}
\includegraphics[width= 1 \textwidth]{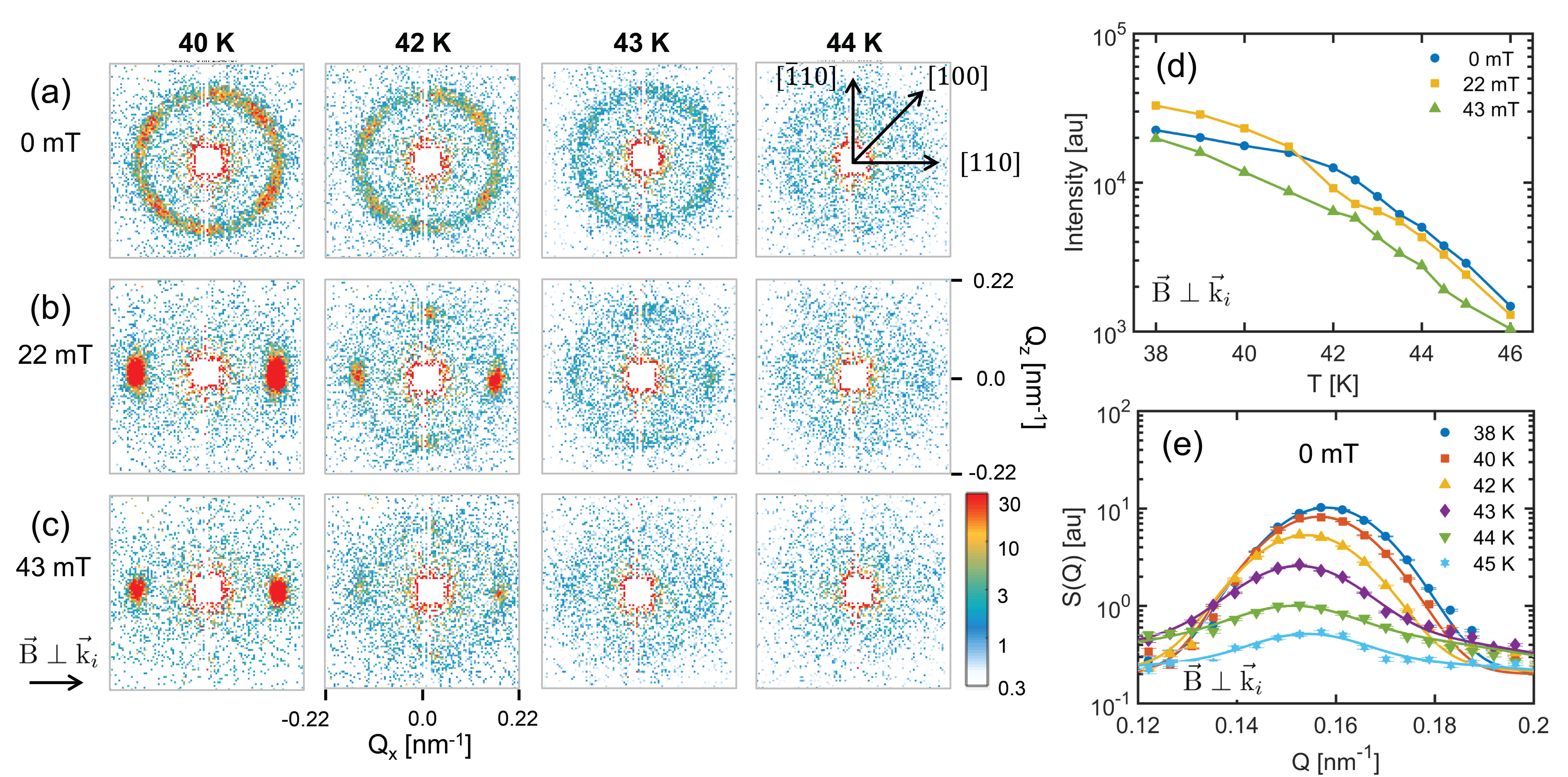}
\caption{SANS results obtained at LARMOR by applying the magnetic field perpendicular to the incident neutron beam ($\vec{B}\perp \vec{k}_i$). Characteristic patterns are shown for $B$ = 0 mT, 22 mT and 43 mT in panels (a) - (c). Panel (d) displays the temperature dependence of the total scattered intensity for selected magnetic fields. Panel (e) shows $S(Q)$ in arbitrary units, deduced by radially averaging the scattered intensity at a magnetic field of 0~mT, for the temperatures indicated. The solid lines indicate the best fits of  eq. \ref{LorGauss} to the data.}
\label{SANS_Larmor}
\end{center}
\end{figure*}

\section{\label{sec:level3} Experimental Results}

\subsection{SANS}
We commence  the presentation of the experimental results with the SANS patterns displayed in Fig. \ref{SANS_D33} for the  configuration where the magnetic field was parallel to the incoming neutron beam ($\vec{B} \, || \,  \vec{k_i}$), and in Fig. \ref{SANS_Larmor} for the complementary set-up where the magnetic field was applied perpendicular to it ($\vec{B} \perp \vec{k}_i$). These results bear the signatures of the different phases present below $T_C$.

At $T$ = 40~K four peaks show up at zero field, that are smeared over a ring with radius $\tau=2\pi/\ell$ and do not have exactly the same intensities due to a slight misalignment of the crystal.\cite{takeda2009, bannenberg2016} These peaks are the signature of the helical phase, where helices are aligned along the $\langle 100 \rangle$ crystallographic directions. By increasing the magnetic field the scattering patterns change and at $B$ = 22~mT there is no scattered intensity for $\vec{B} \parallel \vec{k}_i$, as shown in  Fig. \ref{SANS_D33}(b). On the other hand, only two peaks along the magnetic field direction are found for $\vec{B} \perp \vec{k}_i$ (Fig. \ref{SANS_Larmor}(b)). These patterns are characteristic of the conical phase, where all helices are oriented along the magnetic field and Bragg peaks are thus only visible along the field direction in the configuration where $\vec{B} \perp \vec{k}_i$.

By further increasing the magnetic field a ring of intensity with radius $\tau$ appears. As illustrated by the patterns at  $B$ = 43~mT, this ring is only visible for $\vec{B} \parallel \vec{k}_i$ and as such indicates the onset of the A-phase and skyrmion lattice correlations. These, however, do not lead to the characteristic six-fold pattern found in MnSi due to a combination of magneto-crystalline anisotropy and chemical disorder that is specific to Fe$_{1-x}$Co$_{x}$Si.\cite{munzer2010,bannenberg2016} We note that the skyrmion lattice correlations coexist with the conical phase as the scattering patterns of $\vec{B} \perp \vec{k}_i$ reveal two peaks along the magnetic field direction originating from the conical correlations at the same fields and temperatures for which the skyrmion lattice is stabilized.
  
By increasing the temperature to  $T$ = 42 and 43~K, thus approaching $T_C \approx$ 43~K, the behavior remains roughly the same. However some differences show up as for example the A-phase ring of scattering appears also at $B$ = 22~mT in Fig. \ref{SANS_D33}(b). Furthermore, a broad ring of diffuse scattering is seen at $T$ = 43~K, which resembles the ring of diffuse scattering visible above $T_C$ in MnSi. The patterns in Fig. \ref{SANS_Larmor} indicate that the ring is very weak and coexists with the Bragg peaks of the conical phase and the spots of the A-phase. Above $T$ = 44~K, the intensity of the helical Bragg peaks decreases significantly and are superimposed to a weak ring of diffuse scattering that persists under magnetic field for both configurations.

The temperature dependence of the total scattered intensity, obtained by integrating the SANS patterns, is given for selected magnetic fields in Fig. \ref{SANS_D33}(d) and Fig. \ref{SANS_Larmor}(d)  for $\vec{B} \, || \,  \vec{k_i}$ and  $\vec{B} \perp \vec{k}_i$ respectively. At zero magnetic field the intensity increases gradually with decreasing temperature with no particular change at $T_C$. For $T \gtrsim T_C$ a magnetic field suppresses part of the scattered intensity in both configurations.

Below $T_C$, the temperature dependence is similar for all magnetic fields in the configuration $\vec{B} \perp \vec{k}_i$. On the contrary, for $\vec{B} \, || \,  \vec{k_i}$ a magnetic field has dramatic effects with non-monotonic temperature and magnetic field dependencies. Indeed, even a field of $B$ = 11~mT is large enough to suppress most of the magnetic scattering below $T_C$. Just below $T_C$, diffuse scattering starts to build up at this field, leading to a kink at a temperature that provides the best estimation of $T_C$ from the SANS data. At higher magnetic fields the marked maxima are due to the onset of the A-phase, which appears as additional intensity and is stabilized for 22 $\leq$ $B$ $\leq$ 54~mT. At even higher magnetic fields, as for example for $B$ = 66~mT, the scattered intensity in this configuration is negligible at all temperatures.

\begin{figure}
\begin{center}
\includegraphics[width= .5 \textwidth]{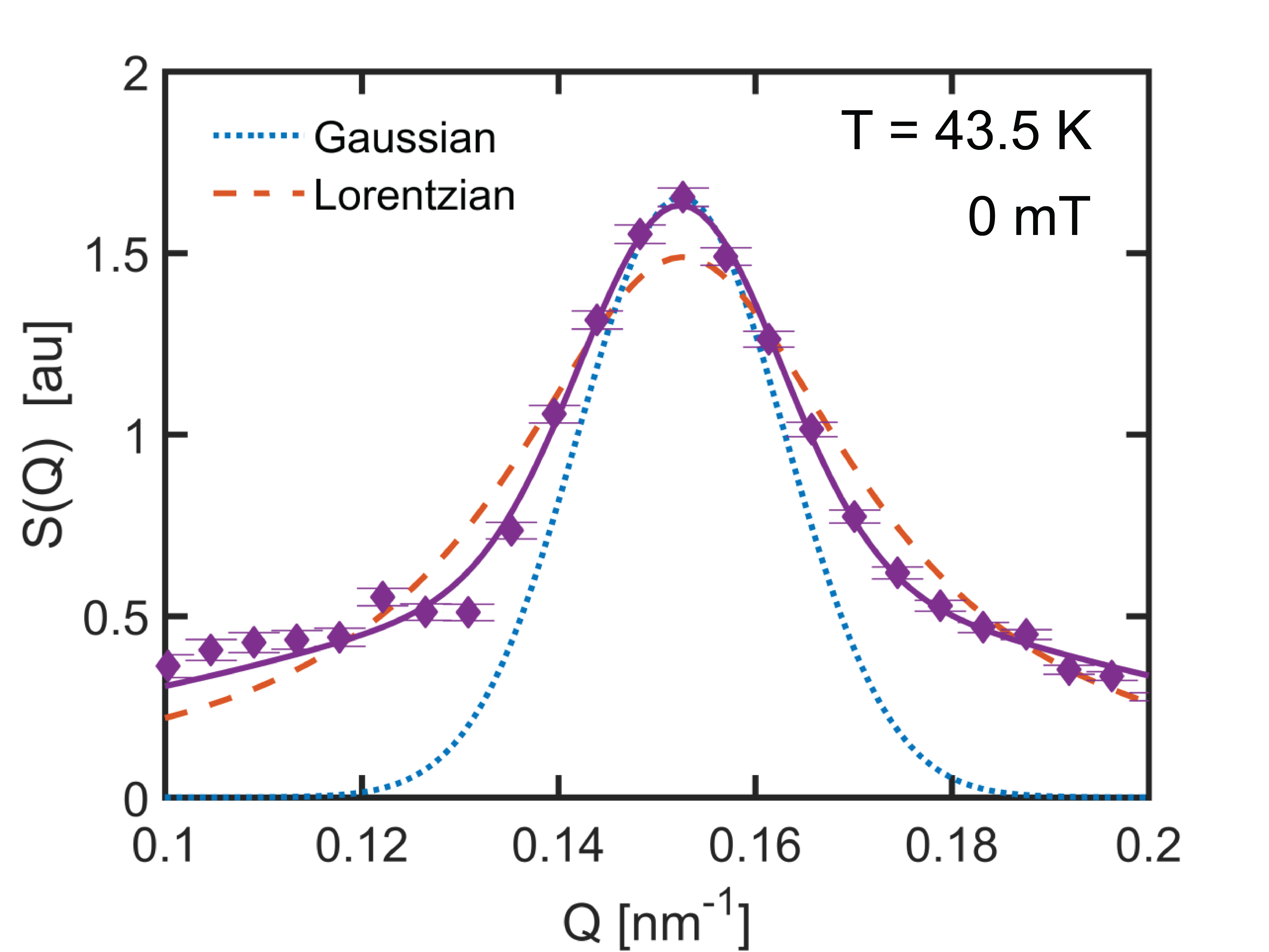}
\caption{Comparison of a Gaussian, Lorentzian and a weighted superposition of a Gaussian and a Lorentzian (solid purple line) to fit $S(Q)$ at $B$ = 0 mT and $T$ = 43.5 K.}
\label{SANS_FitComp}
\end{center}
\end{figure}

\begin{figure*}
\begin{center}
\includegraphics[width= 1 \textwidth]{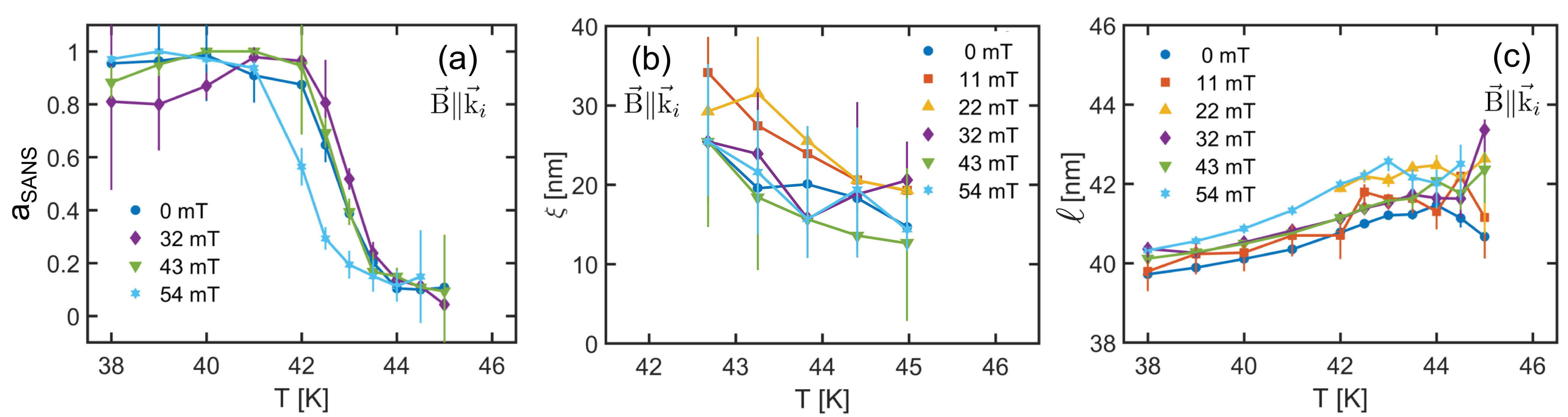}
\caption{The elastic fraction $a_{\text {\tiny \it SANS}}$, the correlation length $\xi$ and the pitch of the helical modulation $\ell$ as a function of temperature for the magnetic fields indicated as derived from the measurements on D33 where the field was applied parallel to the neutron beam ($\vec{B} \parallel  \vec{k}_i$).}
\label{SANS_RadFit}
\end{center}
\end{figure*}

A further step in the quantitative analysis of the SANS data is the analysis of the momentum transfer, $Q$, dependence of the scattered intensity $S(Q)$. $S(Q)$ is obtained in arbitrary units by radial averaging the scattered intensity and is shown in Fig. \ref{SANS_D33}(e) for 43~mT and  Fig. \ref{SANS_Larmor}(e) for 0~mT. Both plots show well defined maxima centered at $\tau = 2\pi/\ell$ and with line-shapes that vary with temperature. In such a plot, Bragg peaks have a Gaussian shape since they are expected to be resolution limited. This contrasts the broad Q-dependence expected for diffuse scattering. In the case of fluctuating correlations with a characteristic correlation length $\xi$ the Ornstein-Zernike formalism predicts the Lorentz function:  
 
\begin{equation}
S(Q) = \frac{C}{(Q-2\pi/\ell)^2+1/\xi^2}, 
\label{LorGauss}
\end{equation}

\noindent with $C$ the Curie constant. Another similar but more complex function has been suggested for chiral magnets,\cite{grigoriev2005} but our experimental results lack the accuracy required to confirm deviations from eq. \ref{LorGauss}. For this reason the data have been analyzed using the simpler Ornstein-Zernike form, convoluted with the corresponding instrumental Gaussian shaped resolution functions. 

At temperatures close to $T_C$, neither a Gaussian nor a Lorentzian function provides a satisfactory description of $S(Q)$ as illustrated in Fig. \ref{SANS_FitComp} for $T$ = 43.5~K. The line shape is satisfactorily reproduced by a weighted superposition of the two functions where the relative weight of the Gaussian function provides an estimate for the elastic fraction $a_{\text {\tiny \it SANS}}$. All $S(Q)$ data have been fitted in this way leading to the values for $a_{\text {\tiny \it SANS}}$, the correlation length $\xi$ and the pitch $\ell$ displayed in Fig. \ref{SANS_RadFit}.

The elastic fraction, displayed in Fig. \ref{SANS_RadFit}(a), is 100\% well below $T_C$, but decreases with increasing temperature above $T$ $\approx$ 41~K and becomes zero within a temperature range of $\sim$ 3~K  at $T \approx$ 44~K.  The deduced values for the correlation length $\xi$ are displayed in Fig. \ref{SANS_RadFit}(b) and show within the experimental accuracy a similar trend for all magnetic fields. At $T$ = 42.5~K,  $\xi$ $\sim$ 30~nm, or $\xi\sim0.75 \cdot \ell$ and decreases with increasing temperature to $\xi\sim 0.35 \cdot \ell$ at 45~K. 

The pitch of the helix $\ell$ shown in Fig. \ref{SANS_RadFit}(c) only depends weakly on the magnetic field. The temperature dependence of $\ell$ is consistent with earlier measurements\cite{bannenberg2016} as $\ell$ increases by $\sim$4\% between $T$ = 38 and 44~K. As the pitch of the helix is proportional to the ratio of the ferromagnetic exchange and the DM interaction, this temperature dependence suggests a slight change in the balance between the two interactions in favor of the ferromagnetic exchange. 

\begin{figure}
\begin{center}
\includegraphics[width= .45 \textwidth]{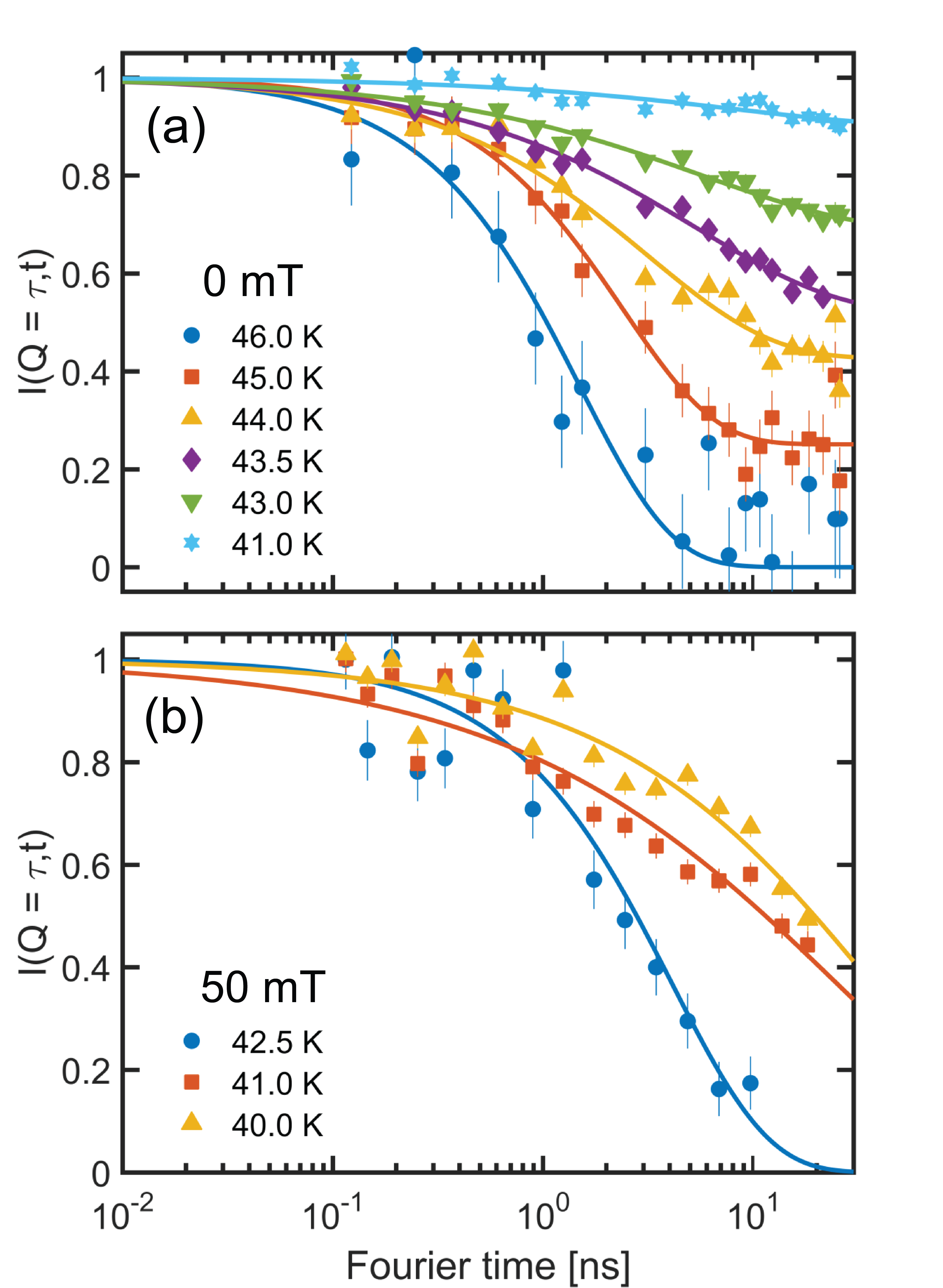}
\caption{Neutron Spin Echo spectroscopy results. Panels (a) and (b) show the intermediate scattering function $I(Q,t)$ measured at (a) 0 mT and (b) 50 mT. The solid lines indicate the fits with the relation provided in eq. \ref{strechedexp}.}
\label{NSESpectra}
\end{center}
\end{figure}

\subsection{NSE}

\begin{figure}
\begin{center}
\includegraphics[width= .5 \textwidth]{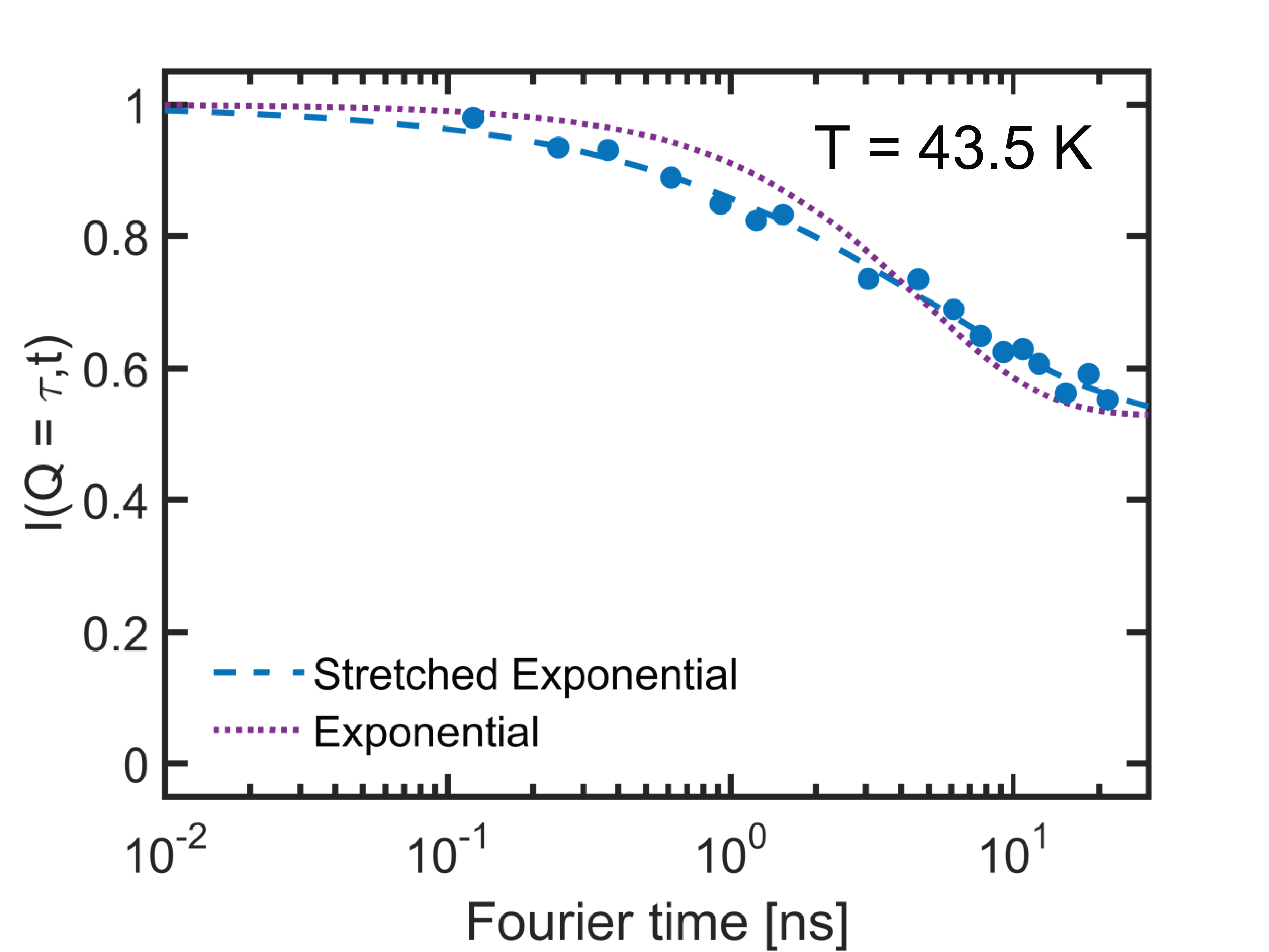}
\caption{Comparison of a fit of the intermediate scattering function $I(Q,t)$  to eq. \ref{strechedexp} exponential ($\beta$ = 1) and a stretched exponential fit (0 $\leq$ $\beta$ $\leq$ 1) for $B$ = 0 mT and $T$ = 43.5 K.}
\label{NSE_FitComp}
\end{center}
\end{figure}

\begin{figure*}
\begin{center}
\includegraphics[width= 1 \textwidth]{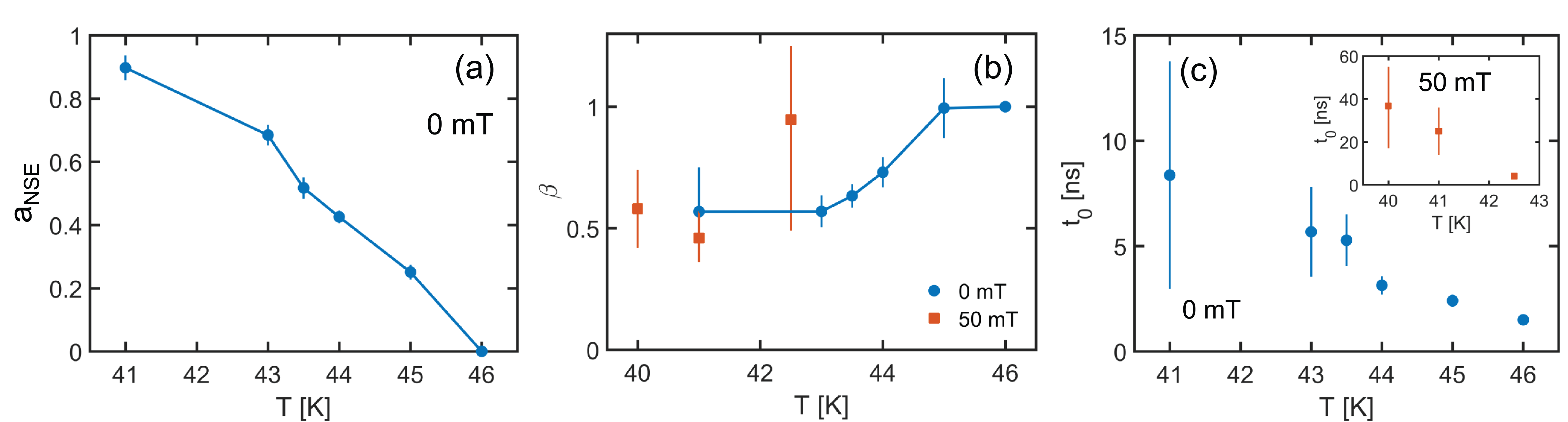}
\caption{Obtained parameters from fitting the intermediate scattering function $I(Q,t)$ with the relation provided in eq. \ref{strechedexp}. Panel (a) shows the elastic fraction $a_{\text {\tiny \it NSE}}$, panel (b) the relaxation time $t_0$ and panel (c) the stretching exponent $\beta$ as a function of temperature at $B$ = 0 mT and $B$ = 50 mT.}
\label{NSE_Fit}
\end{center}
\end{figure*}

The relaxation of the magnetic correlations above $T_C$ has been investigated by NSE at $B=$ 0 and 50~mT. The corresponding intermediate scattering functions $I(Q,t)$ are displayed as a function of the Fourier time $t$ in Fig. \ref{NSESpectra}(a) and (b), respectively. Above $T_C$, at 46~K for $B$ = 0~mT and 42.5~K for  $B$ = 50~mT, the relaxation is exponential and  $I(Q,t)$  decays from 1 to 0 when $t$  increases by less than two orders of magnitude. However, when the temperature decreases and approaches $T_C$, the relaxation stretches and covers a much larger time domain. In addition, $I(Q,t)$ at zero field levels off at the longest Fourier times to an elastic component, $a_{\text {\tiny \it NSE}}$,  that increases with decreasing temperature. The time decay of $I(Q,t)$  has therefore been fitted by a superposition of the elastic component and a stretched exponential relaxation: 

\begin{equation}
I(Q,t)=\left(1-a_{\text {\tiny \it NSE}} \right)\exp\left[-\left({t}/{t_0}\right)^\beta\right]+a_{\text {\tiny \it NSE}}
\label{strechedexp}
\end{equation}

\noindent  with $t_0$ the characteristic relaxation time and $\beta$ the stretching exponent. The necessity to include the stretching exponent $\beta$ is illustrated in Fig. \ref{NSE_FitComp}, which displays the time decay of $I(Q,t)$ at $T$ = 43.5~K and shows that the fit with a simple exponential ($\beta$ = 1) is poor. The decay is more stretched than an exponential and leads in this case to $\beta$ $\sim$ 0.6.

The temperature dependence of the parameters deduced from the fit of the NSE spectra at zero field is given in Fig. \ref{NSE_Fit}. At $T$ = 46~K, $a_{\text {\tiny \it NSE}}\sim$ 0, but the elastic fraction increases roughly linearly with decreasing temperature reaching $\sim$ 90\% at $T$ = 41~K, which is in qualitative agreement with the values obtained from the SANS measurements $a_{\text {\tiny \it SANS}}$ displayed in Fig. \ref{SANS_RadFit}(a). The stretching exponent $\beta$ depicted in Fig. \ref{NSE_Fit} (b) is equal to one well above 45~K but decreases with decreasing temperature and reaches $\beta$ $\sim$ 0.57 at 41~K. The characteristic relaxation times $t_0$, displayed in Fig. \ref{NSE_Fit}(c), increase gradually with decreasing temperature from $t_0$ = 1.7~ns at 46~K to $t_0$ = 8~ns at $T$ = 41~K, indicating a slowing down of the relaxation with decreasing temperature approaching $T_C$.

The data at $B$ = 50~mT are less precise than at zero field, as they have been collected in the ferromagnetic NSE configuration that is sensitive to background corrections. The spectra do not allow a reliable determination of $a_{\text {\tiny \it NSE}}$ and for this reason no data are given in Fig. \ref{NSE_Fit} (a). On the other hand, the stretching exponent $\beta$ shows a similar behavior as at zero field. In contrast to $\beta$, the characteristic relaxation times are almost doubled as compared with zero field and reach $t_0$ $\sim$40~ns at $T$ = 40~K. As such, these longer relaxation times indicate a considerable slowing down of the dynamics under field.

\subsection{Magnetic Chirality}

Following the equations of Blume and Maleyev,\cite{Blume, Maleyev2} the interaction between a polarized neutron beam and chiral magnetic correlations affects the polarization and the intensity of the scattered neutron beam. This interaction reveals both the handness $\zeta$, with $\zeta=+1$ for right and $=-1$ for left handed chirality, respectively, and the weight $\eta(\vec{Q})$ of the dominant chiral domain, with  $\eta$ = 1 for a single chirality domain and $\eta$ = 0 for a disordered state or domains with equally populated chiralities. For a perfectly polarized incident neutron beam, this can be written as:

\begin{equation}
\zeta \eta(\vec{Q})=|\vec{M}(\vec{Q})_\perp \times \vec{M}(\vec{Q})_\perp^\ast|/ (\vec{M}(\vec{Q})_\perp \cdot \vec{M}(\vec{Q})_\perp^\ast)
\label{chiral}
\end{equation}

\noindent  with $\vec{M}(\vec{Q})_\perp$ the projection of the magnetic structure factor onto a plane perpendicular to the scattering vector $\vec{Q}$. This term can be determined independently from the polarization and the intensity of the scattered beam.\cite{pappas2009,pappas2011}

With spherical neutron polarimetry it is possible to control the incident and the scattered neutron beam polarization vectors,  $\vec{P}'$ and  $\vec{P}$ respectively, independently from each other. The measurements determine  the polarization transfer matrix, with respect to a right-handed Cartesian set with $\hat{x}$ = $\hat{Q}$, $\hat{z}$ in the scattering plane and $\hat{y}$ perpendicular to it.\cite{BrownPNCMI2000} In the small angle scattering geometry used in this experiment $\hat{y}$ was collinear with $\vec{k_i}$. Table \ref{table} provides the ideal matrix for the case of a chiral helix and the experimental results for Fe$_{0.7}$Co$_{0.3}$Si at 42~K revealing full right handed chirality. The error bars printed in the table are deduced from the counting rates and do not include systematic errors of the polarimetric setup or the sample alignment, which likely account for the slight discrepancies between the measured and expected values. 
 
\begin{table}[htp]
\setlength{\tabcolsep}{2.5mm}
\renewcommand{\arraystretch}{1.2}
\caption{Ideal and measured polarization matrix $\mathbb{P}_{\alpha,\beta}$ at  $\vec{\tau}_{100}$. }
\begin{center}
\begin{tabular}{|c@{}|r@{}r@{}r@{}|r@{.}lr@{.}lr@{.}l|}
\hline
 \multicolumn{1}{|r|}{$ $} &   \multicolumn{3}{c|}{ideal} & \multicolumn{6}{c|}{T = 42 K}  \\
\hline
$\beta\; \diagdown\, \alpha$$\;\;$ & \multicolumn{1}{r}{$\;\;x$} & \multicolumn{1}{c}{$\;\; y$} & \multicolumn{1}{c|}{$\,\,  z$} & \multicolumn{2}{c}{$x$}& \multicolumn{2}{c}{$y$} & \multicolumn{2}{c|}{$z$}  \\
\hline

$x$ &  -1    &	$\eta\,\zeta$	& $ \eta\,\zeta$ $ \;$ 	& -0&92(3) & 1&03(5) & 0&99(5) \\
$y$ &   0    & 	0   			& 0 $\;$			 &  0&10(2) &  0&22(5) &   -0&15(4)\\
$z$ &   0    & 	0   			& 0 $\;$			 & 0&07(2) & -0&10(4) & 0&06(4) \\
\hline
\end{tabular}
\end{center}
\label{table}
\end{table}%

Complementary results are obtained from the difference in the scattered intensities, in spin flip configuration, when the incoming beam polarization is parallel $N^{\text{\it \tiny SF}}_{\vec{\text{\it \tiny P'}}\parallel {\vec{\text{\it \tiny Q}}}}=N_{\hat{x}, -\hat{x}}$ or anti-parallel $N^{\text{\it \tiny SF}}_{\vec{\text{\it \tiny P'}}\parallel -{\vec{\text{\it \tiny Q}}}}=N_{-\hat{x}, \hat{x}}$ to $\vec{Q}$. This difference is zero in the absence of a chiral term and  after background correction the product $\zeta \eta(\vec{Q})$ is obtained directly from the ratio $\zeta \eta(\vec{Q}) =(N_{-\hat{x},\hat{x}} - N_{\hat{x},-\hat{x}} )/(N_{\hat{x}, -\hat{x}}+N_{-\hat{x}, \hat{x}})$.\citep{pappas2009,pappas2011}

\begin{figure}
\begin{center}
\includegraphics[width= 0.45 \textwidth]{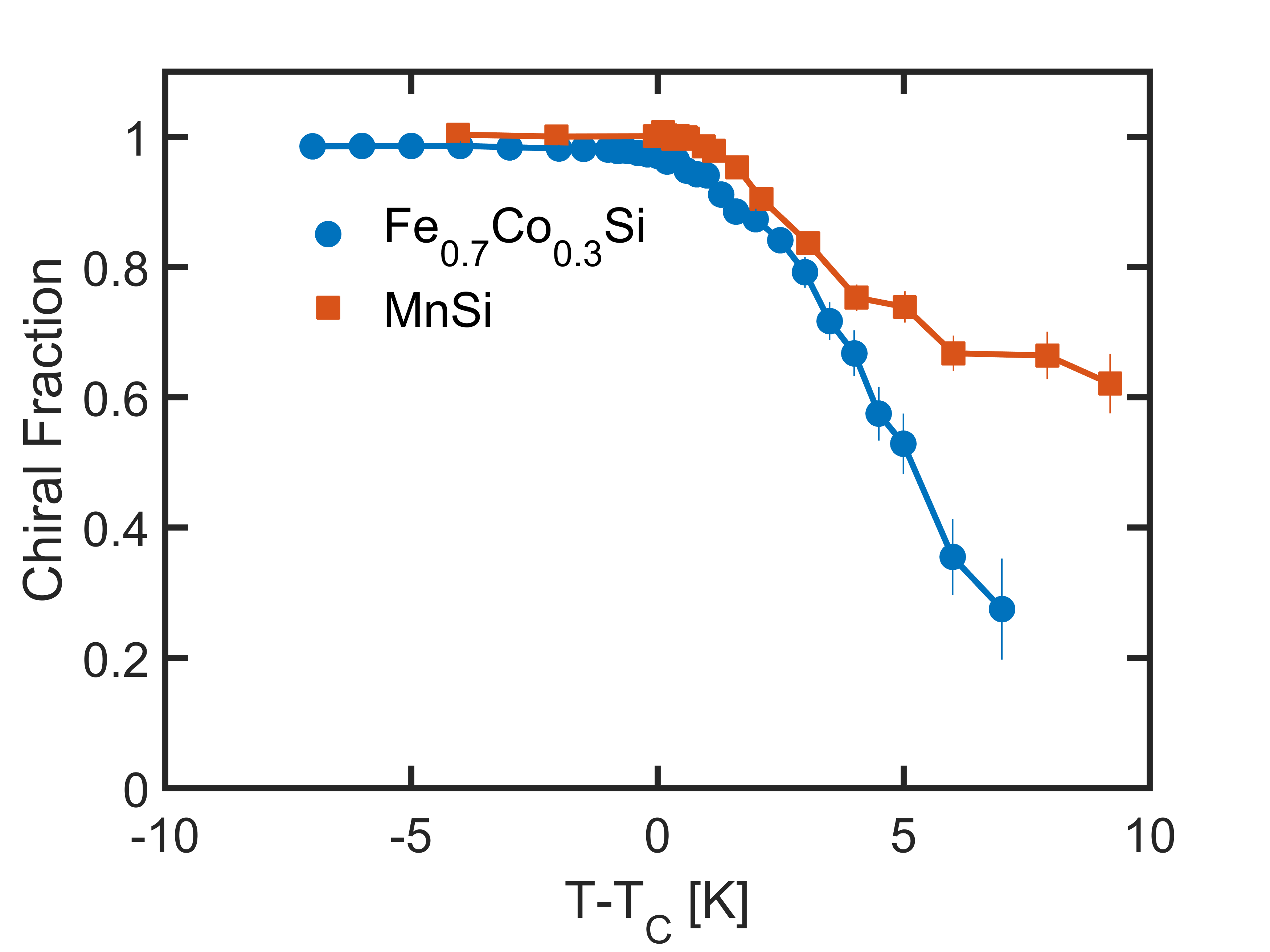}
\caption{The chiral fraction as a function of $T-T_C$ for both Fe$_{0.7}$Co$_{0.3}$Si and MnSi at $B$ = 0 mT. The data from MnSi have been adapted from refs.\citep{pappas2009,pappas2011}}
\label{NSE_Chirality}
\end{center}
\end{figure}


The resulting temperature dependence of $\eta(\vec{Q}=\vec{\tau})$ after correction for the incident beam polarization is displayed in Fig. \ref{NSE_Chirality}, where also the corresponding chiral fraction for MnSi is included as a reference. It shows that at $\vec{Q}=\vec{\tau}$, $\eta$ = 1 below $T_C \sim$ 43~K, implying that the system is completely chiral below $T_C$. Above $T_C$, $\eta$ decreases substantially with increasing temperature and drops to $\eta$ = 0.80 at $T$ = 46~K. However, $\eta$ remains non-zero even at 50~K, which implies that well above $T_C$ the system is still influenced by DM interactions.

\section{Discussion}
In the following we will discuss our experimental findings in the context of the literature, and in particular in comparison with the archetype chiral magnet MnSi. As a first step, we determine the characteristic length scales relevant to the transition, discuss their hierarchy and the applicability of the Brazovskii approach suggested for MnSi.\citep{janoschek2013} Subsequently, we directly compare the helimagnetic transitions in Fe$_{0.7}$Co$_{0.3}$Si and MnSi and highlight the particularities of Fe$_{0.7}$Co$_{0.3}$Si.

\subsection{\label{sec:level6} Characteristic lengths and the Brazovskii Approach}
\begin{figure}
\begin{center}
\includegraphics[width= .45 \textwidth]{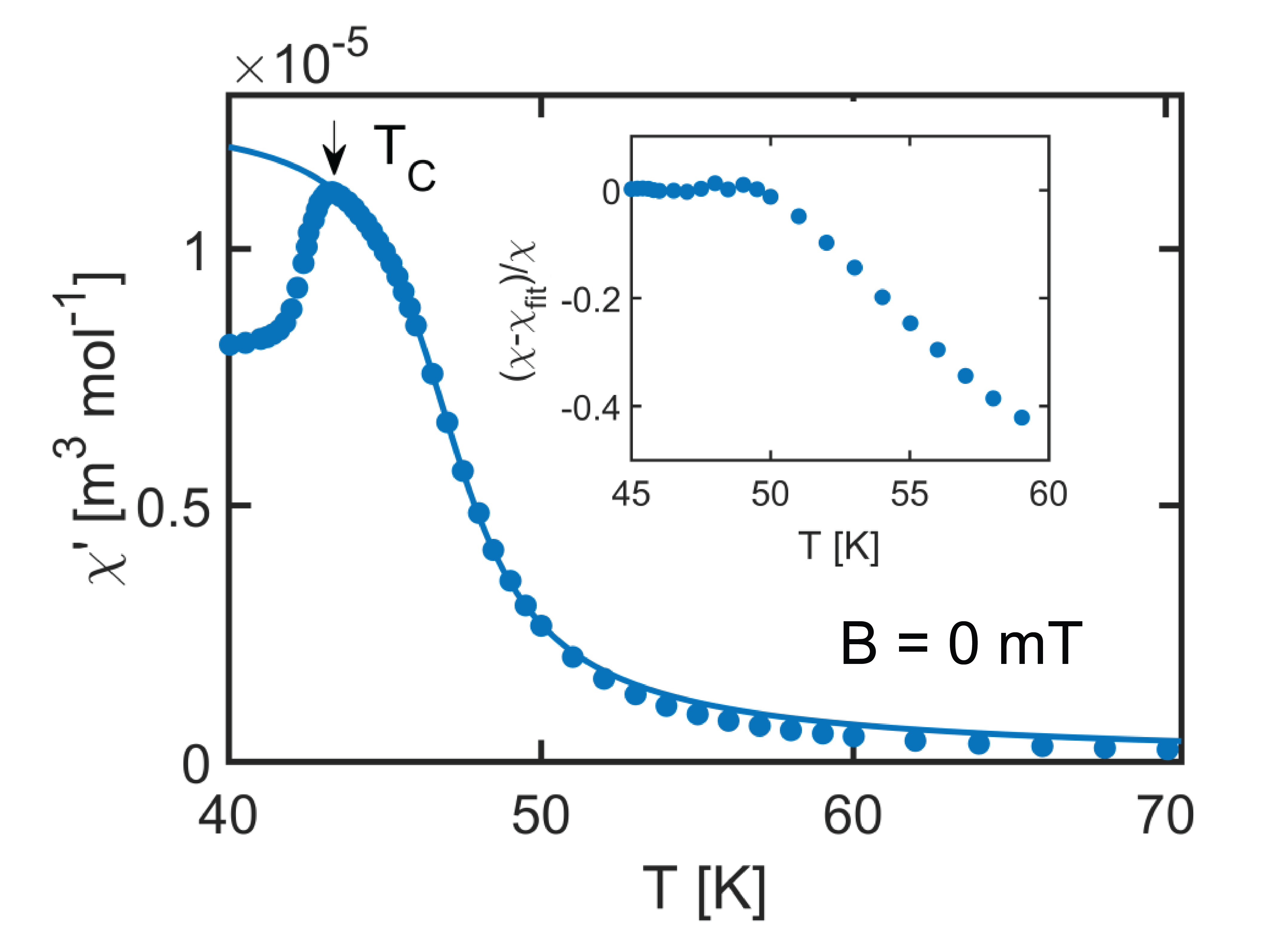}
\caption{Zero field ac susceptibility as a function of temperature measured at $f$ = 5~Hz and reproduced from ref.\cite{bannenberg2016squid}. The solid line indicates the fit to the relation of eq. \ref{braz1}. The inset shows the relative difference of the measured data to the fit with eq. \ref{braz1}.}
\label{chi}
\end{center}
\end{figure}

It has been suggested that the helimagnetic transition is governed by a hierarchy of characteristic lengths reflecting the relative strength of the interactions:\cite{bak1980,janoschek2013} the Ginzburg length $\xi_{G}$, the Dzyaloshinsky-Moriya (DM) length $\xi_{DM}$, and $\xi_{cub}$, the length associated with the cubic anisotropy. The Ginzburg length quantifies the strength of the interactions between the magnetic fluctuations, which are strong in the limit of $\xi_{G}$ $>$ $\xi(T)$, whereas they can be considered as a perturbation in the other limit $\xi(T)$ $<$ $\xi_G$. The Dzyaloshinsky-Moriya (DM) length is related to the pitch of the helix $\ell$ by $\xi_{DM}$ = $\ell$/2$\pi$ and the cubic anisotropy length reflects the influence of the cubic anisotropy.

If the Ginzburg length is much larger than the length scale associated with the DM interaction, i.e. $\xi_{G}$ $>>$ $\xi_{DM}$, the interactions between the fluctuations should govern the behavior close to the transition, driving it to first order as suggested by Brazovskii.\cite{brazovskii1975,janoschek2013} Also for $\xi_{G}$ $<$ $\xi_{DM}$ the transition is driven by fluctuations and expected to be of first order following the Bak and Jensen (1980) approach in this so called Wilson-Fisher renormalization group limit. So far, the helimagnetic transition and the role of the characteristic lengths remain largely unexplored and have only been discussed for the case of MnSi\cite{janoschek2013} and Cu$_2$OSeO$_3$.\cite{zivkovic2014}

Whereas the DM length can be obtained directly from the pitch of the helix, the cubic anisotropy length can only be derived from the temperature dependence of the correlation length. On the other hand, the Ginzburg length can be obtained both from the correlation length and the macroscopic susceptibility. As the accuracy of the experimentally determined correlation length is limited, we are unable to determine the cubic anisotropy length and we derive the Ginzburg length from the susceptibility data of ref.\cite{bannenberg2016squid} reproduced in Fig. \ref{chi}. For this purpose, we fit the real component of the ac susceptibility $\chi^\prime$ with:\cite{brazovskii1975,janoschek2013}

\begin{equation}
\chi^\prime = \frac{\chi^\prime_0}{1+\sigma^2Z^2(T)},
\label{braz1}
\end{equation}

\noindent  where $\chi_0$ is a constant, $\sigma$ equals the ratio of the DM and Ginzburg length ($\sigma$ = $\xi_{DM}$/$\xi_{G}$) and $Z(T)$ is given by:

\begin{equation}
Z(T) = \frac{\sqrt[3]{2}\epsilon+(1+\sqrt{1-2\epsilon^3})^{2/3}}{\sqrt[3]{2}(1+\sqrt{1-2\epsilon^3})^{1/3}},
\label{braz2}
\end{equation}

\noindent  where $\epsilon=(T-T_{MF})/T_0$ is a relative measure of the distance to the mean field temperature $T_{MF}$. 

Equation \ref{braz1} provides a good description of the temperature dependence of $\chi^\prime$ up to $T \approx$ 50~K as shown in Fig. \ref{chi}, leading to estimates of $\chi_0$ = 1.27 $\pm$ 0.03  10$^{-5}$ m$^3$ mol$^{-1}$, $T_{MF}$ = 48.3 $\pm$ 0.2 K, $T_0$ = 3.2 $\pm$ 0.2 K and $\sigma$ = 1.39 $\pm$ 0.08. As illustrated by the inset of Fig. \ref{chi}, substantial and systematic deviations occur above 50~K. The value of $\xi_{DM}$ $\approx$ 6.3~nm derived from the neutron data (Fig. \ref{SANS_RadFit}(c)) translates $\sigma$ to $\xi_{G}$ $\sim$ 4.5~nm. Both lengths are thus much shorter than the correlation lengths, which exceed 10~nm in the probed temperature range below 45~K (see Fig. \ref{SANS_RadFit}(b)). 

As $\sigma$ $>$ 1, and therefore $\xi_{G}$ $<$ $\xi_{DM}$, the applicability of the Brazovskii approach for Fe$_{0.7}$Co$_{0.3}$Si is questionable. The obtained value for $\sigma$ of 1.39 is much larger than the value found for MnSi\citep{janoschek2013} ($\sigma \approx$ 0.50) but is in fact reasonably similar to the value of $\sigma$ = 1.18(1) found for Cu$_2$OSeO$_3$.\citep{zivkovic2014} The comparison of the characteristic lengths therefore indicates that the helimagnetic transition in Fe$_{0.7}$Co$_{0.3}$Si should be different from that in MnSi.

\subsection{Comparison with MnSi and other cubic helimagnets}

\begin{figure}
\begin{center}
\includegraphics[width= 0.45 \textwidth]{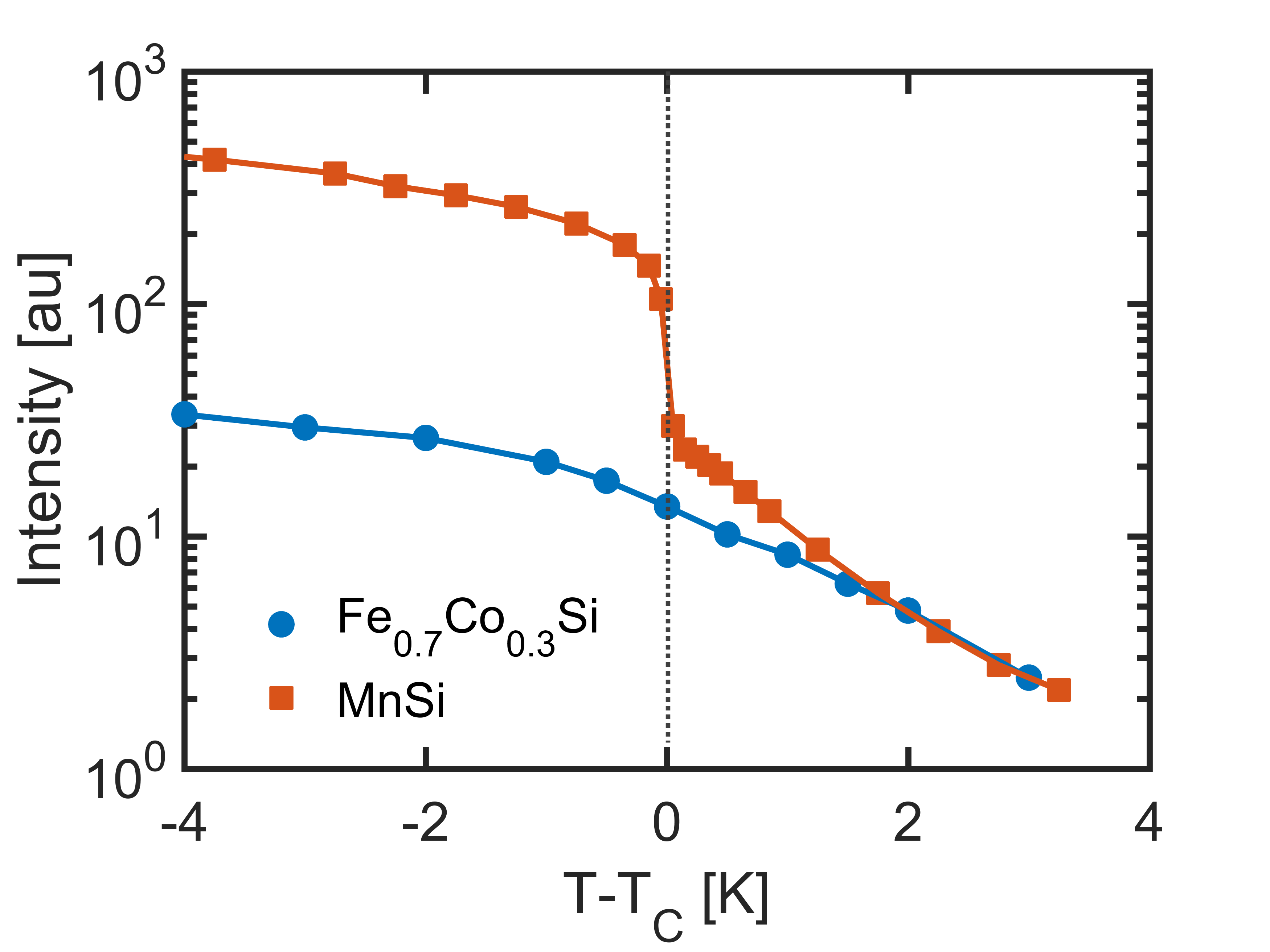}
\caption{Total background corrected scattered intensity obtained by summing the entire detector as a function of temperature at $B$ = 0~mT for Fe$_{0.7}$Co$_{0.3}$Si and MnSi as measured on LARMOR. MnSi data adapted from \cite{pappas2016} and scaled at the highest temperature indicated.}
\label{Comp}
\end{center}
\end{figure}

The difference between Fe$_{0.7}$Co$_{0.3}$Si and MnSi is highlighted by comparing the total scattered intensity at zero field as displayed in Fig. \ref{Comp}. The total intensity is obtained by summing the intensity over the entire detector and thus also outside the Bragg peak positions. At high temperatures a similar temperature dependence is found for both chiral magnets. Differences in intensity start to develop at $T_C$ + 2 K although the chiral behavior is similar in this region (Fig. \ref{NSE_Chirality}). In MnSi, an increase of intensity indicates the  onset of the precursor phase and strong buildup of chiral correlations, which ultimately results in a jump of scattered intensity by more than an order of magnitude within a narrow region of 0.2~K around $T_C$. This jump is as such a signature of the first-order nature of the transition. In contrast to MnSi, in Fe$_{0.7}$Co$_{0.3}$Si the increase of scattered intensity with decreasing temperature is gradual and without strong indication for a precursor phase and a jump in intensity at $T_C$. 

The onset of the helimagnetic order is thus very different for the two chiral magnets. This is in agreement with specific heat measurements as the sharp peak visible at $T_C$ for MnSi\cite{bauer2013} or Cu$_2$OSeO$_3$\citep{adams2012} has not been seen for Fe$_{0.7}$Co$_{0.3}$Si.\cite{bauer2016} The gradual transition manifests itself also by the coexistence of long-range helimagnetic order and short-range chiral correlations. This coexistence occurs in Fe$_{0.7}$Co$_{0.3}$Si over a wide temperature range of $\sim$5~K, whereas in MnSi this region does not exceed $\sim$0.4~K.\citep{pappas2009,pappas2011}  

In addition, Fig. \ref{Comp} shows that the build-up of correlations above $T_C$ in the precursor region is not as strong as for MnSi. In the later, intensive correlations start to build up in this region of the phase diagram. This is also the case for Fe$_{0.7}$Co$_{0.3}$Si but the probed correlations are much weaker, possibly because the precursor phenomena are partially suppressed by  substitutional disorder. The precursor phase and the associated precursor phenomena in cubic chiral magnets have been subject to intensive theoretical and experimental studies. \cite{rossler2006, wilhelm2011, wilhelm2012,moskvin2013} It has been suggested that a softening of the magnetization plays an important role in the formation of isolated skyrmions and other localized magnetic states in this region of the phase diagram.\cite{rossler2006, wilhelm2011, wilhelm2012, moskvin2013, grigoriev2014} 

The dynamics of the helimagnetic transition as probed by NSE also reveal major differences between the two chiral magnets. Whereas the dynamics of MnSi can be described by a simple exponential, the dynamics of Fe$_{0.7}$Co$_{0.3}$Si around $T_C$ can only be accounted for by a stretched exponential relaxation.\cite{pappas2009,pappas2016} This implies that the relaxation stretches over several orders of magnitude in time and involves a broad distribution of relaxation times. The corresponding stretching exponents are close to the values found in disordered systems such as glass forming systems.\cite{Mezei:1987uf, colmenero2003} In addition, the associated characteristic relaxation times at zero field for Fe$_{0.7}$Co$_{0.3}$Si can be as long as 10~ns, and are therefore much longer than for MnSi where they do not exceed 1~ns.\citep{pappas2009,pappas2016} These long relaxation times combined with the stretched exponential relaxation suggest the relaxation of large magnetic volumes, that are likely inhomogeneous in size and structure. Similar conclusions have also been drawn from ac susceptibility measurements below $T_C$, which however probe much longer, macroscopic, relaxation times.\citep{bannenberg2016squid} 

The SANS and NSE results show that the helical transition in Fe$_{0.7}$Co$_{0.3}$Si is gradual and involves more complicated and slower dynamics, than in MnSi. The hierarchy of the characteristic length scales puts Fe$_{0.7}$Co$_{0.3}$Si closer to Cu$_2$OSeO$_3$ than to MnSi, but this hierarchy in itself does not explain the particularities of the helimagnetic transition in Fe$_{0.7}$Co$_{0.3}$Si. Moreover, neither the Brazovskii nor the Wilson Fisher approach can describe the transition in Fe$_{1-x}$Co$_x$Si as both approaches predict a sharp first order helimagnetic transition that is in both cases driven by fluctuations.\citep{brazovskii1975,bak1980,janoschek2013} 

One important factor that could possibly explain the different transition in Fe$_{1-x}$Co$_x$Si is the chemical disorder that arises from the solid solution of Fe and Co in Fe$_{0.7}$Co$_{0.3}$Si. In conjunction with the effect of the cobalt concentration on the sign of the DM-interaction, the chemical disorder might be an additional source of frustration. This chemical disorder could effectively `smear' the phase transition and make it appear more continuous. This would explain the broad distribution of relaxation times and the stretched exponential relaxation probed by NSE. However, we note that the stretched exponential relaxation does not reflect the inhomogeneous relaxation of regions with different chiralities as the magnetic chirality is 100\% close to $T_C$ (Fig. \ref{NSE_Chirality}).

\subsection{Transition under magnetic field}
The SANS results show that the transition under field remains similar to the one at zero field up to 40~mT. For these relatively low magnetic fields, the scattered intensity above $T_C$ is similar to the intensity at zero field. Furthermore both the correlation lengths and the elastic fractions remain almost unchanged.  For fields exceeding 40~mT, the scattered intensity decreases until it is almost completely suppressed for fields above 60~mT. This is highly similar to the behavior of the transition in MnSi, where the scattered intensity is also suppressed by large enough magnetic fields.\citep{pappas2016}

In addition, the dynamics of the transition to the conical phase, probed by NSE, slows down considerably under magnetic field. Compared to zero field, the relaxation times under field almost quadruple and can be as long as 40~ns at $B$ = 50~mT. Although the magnetic field induces much sharper peaks in the conical phase than at zero field (Fig. \ref{SANS_Larmor}, the relaxation remains strongly non-exponential. Therefore, the broad distribution of relaxation times is not affected by the magnetic field.

\section{\label{sec:level5} Summary and Conclusion}
In conclusion, SANS and NSE reveal a very gradual and smeared transition around $T_C$ that differs substantially from the sharp first order phase transition in MnSi. Magnetic correlations that are partially chiral but much weaker than in MnSi coexist in a wide temperature range with the long-range helimagnetic order. The relaxation around $T_C$ is broad, non-exponential, even under magnetic field, and is much slower and more complex than in MnSi. The hierarchy of interactions and of the deduced length scales places Fe$_{0.7}$Co$_{0.3}$Si closer to Cu$_2$OSeO$_3$ than to MnSi, but cannot explain all the particularities of the helimagnetic transition. The large differences between the transition in Fe$_{1-x}$Co$_x$Si and other systems of the same family challenges the validity of an universal approach to the helimagnetic transition in chiral magnets.

\begin{acknowledgments}
The authors wish to thank the ISIS and ILL technical support staff for their assistance. The work of LB is financially supported by The Netherlands Organization for Scientific Research through project 721.012.102. FQ acknowledges support from the Chinese Scholarship Council and the European Union Seventh Framework Programme [FP7/2007-2013] under grant agreement Nr. 283883.
\end{acknowledgments}

\bibliography{FeCoSi_Tc_Paper}

\end{document}